# Design of Mutation Operators for Testing Geographic Information Systems [†]

**Suilen H. Alvarado**

Laboratorio de Bases de Datos Campus de Elviña, Centro de investigación CITIC, Universidade da Coruña, 15071 A Coruña, Spain; s.hernandez@udc.es

† Presented at the 2nd XoveTIC Congress, A Coruña, Spain, 5–6 September 2019.

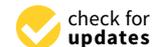

**Abstract:** In this article, we propose the definition of specific mutation operators for testing Geographic Information Systems. We describe the process for applying the operators and generating mutants, and present a case study where these mutation operators are applied to two real-world applications.

## 1. Introduction

Mutation-based testing [1] is a test technique that involves artificially introducing errors into a System Under Test (SUT). A *mutant* is a copy of the system in which a change has been done that, in most cases, will lead to a behaviour different than expected. The different mutants are generated automatically by the application of mutation operators.

In the state of the art, we have found mutation operators, both general purpose and specific to different technologies, languages and paradigms [2–9]. However, these operators are not adequate when trying to test software features associated with specific domains.

In this article, we propose mutation operators specific to the domain of Geographic Information Systems (GIS) applications. These operators reproduce programming errors that are litely to occur during the development of this type of applications. In addition, we present the implementation of these operators and as proof of concept we apply these operators to two real-world GIS applications and we generate the mutants.

## 2. Mutation Operators for GIS

As a previous step to designing the mutation operators, we analyzed the main technologies used specifically in the development of GIS, and we identified typical errors a programmer can introduce during the development. These errors were formalized into mutation operators. In order to apply these operators to a SUT, we rely on Java reflection and aspect-oriented programming. Reflection allows us to obtain the list of classes and methods of the SUT, so the user can decide the methods to wish the operators will be applied.

Later, we capture information about the methods of the SUT to be mutated, together with the information of the mutation operators that were already defined. From these data, we generate the mutation operator, in the form of on aspect, which will then be possible to interweave with the SUT which generates a mutant of the SUT.

Next, we describe the definition of two operators and two cases of application on real-world GIS applications.

**ChangeCoordSys Operator (Listing 1):** It exchanges the coordinate system of a geometry, so it does not match the coordinate system that is being used in the user interface. It simulates the error of not checking that the coordinate system is correct. The error is introduced by directly modifying the coordinate system of geometry when recovering the wrapping of the figure.

```
    public String getCode(String code) {
    code="double temp = (double) args[0];
            args[0] = (double) args[1];
            args[1] = temp;
            ";
    return code;
    }

    public String[] getOperationsNames() {
  return new String[] { "getFromLocation" };
    }
}
```

Listing 1: A simplified definition of the ChangeCoordSys Operator.

This operator was applied to a mobile technology GIS application. This application allows registering places of interest for the user. These areas of interest are called *Geofences*. A Geofence is determined by a geographical location expressed in terms of latitude, longitude, and a radius around that location. By creating a Geofence with an erroneous location from its central location, the device will receive incorrect location notifications. As a result, the user will see in the application's map viewer the Geofences drawn in erroneous zones (Figure 1).

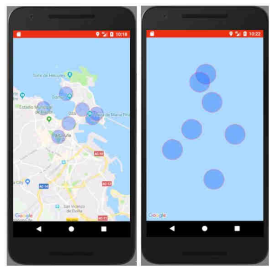

**Figure 1.** Original and mutant application.

**BooleanPolygonConstraint Operator (Listing 2):** It introduces errors in the processing of geometries, manipulating the result of the operations that carry out the verification of different topological restrictions between geometries, such as intersects, covers or overlap.

```
    public String getCode(String code) {
      code = "com.vividsolutions.jts.geom.Coordinate[]
      coordinates = pGeometry1.getCoordinates();\n" +
      "coordinates[0] = pGeometry1.getCentroid().getCoordinate();\n" +
      "coordinates[coordinates.length-1] =
       pGeometry1.getCentroid().getCoordinate();\n" +
      "pGeometry1 = new com.vividsolutions.jts.geom.GeometryFactory()
      .createPolygon(coordinates);\n" +
      "args[0] = pGeometry1;";
      return code;
    }

    protected String[] getOperationsNames() {
  return new String[] {"contains", "coveredBy", "covers", "crosses",
  "disjoint", "touches", "equalsTop", "intersects",
  "overlaps", "within"};
    }
}
```

Listing 2: A simplified definition of the BooleanPolygonConstraint Operator.

To test this operator it was applied to a land reparcelling system. The objective of the land reparcelling is to reunify the lands of an owner to facilitate their exploitation. In this application, the result of the operation between two polygons has been affected. This error causes the incorrect display of the resulting geometry that should be drawn in the user interface after the operation applied to the two initial geometries (Figure 2).

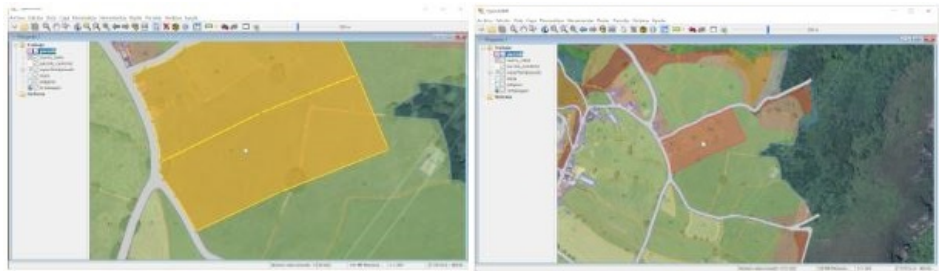

**Figure 2.** Original and mutant application.

## 3. Conclusions

In existing proposals, we can find both generic and specific mutation operators. However, these are not adequate to cover errors in particular domains. We have defined new operators specific to the GIS domain and a way to apply them to a SUT. In addition, we have tested the operators defined in two GIS applications. As future work, we intend to extend this approach to other domains, as well as to use the developed operators for the automatic improvement of sets of test cases.

**Funding:** Financed by: Xunta de Galicia / FEDER-UE CSI: ED431G/01 (Centros singulares de investigación de Galicia), Xunta de Galicia / FEDER-UE CSI: ED431C 2017/58 (Grupo de Referencia Competitiva), MINECO-AEI/FEDER-UE: Datos 4.0 (TIN2016-78011-c4-1-R) and BIZDEVOPS-GLOBAL (RTI2018-098309-B-C32), and EU H2020 MSCA RISE BIRDS: 690941 (S.H.A.).